# PROTECTING ORACLE PL/SQL SOURCE CODE FROM A DBA USER

Hakik Paci, Elinda Kajo Mece, Aleksander Xhuvani[1]

[1]Polytechnic University of Tirana, Sheshi Nene Tereza, Tirane, Albania
hpaci@fti.edu.al, ekajo@fti.edu.al, axhuvani@fti.edu.al

## ABSTRACT

*In this paper we are presenting a new way to disable DDL statements on some specific PL/SQL procedures to a dba user in the Oracle database. Nowadays dba users have access to a lot of data and source code even if they do not have legal permissions to see or modify them. With this method we can disable the ability to execute DDL and DML statements on some specific pl/sql procedures from every Oracle database user even if it has a dba role. Oracle gives to developer the possibility to wrap the pl/sql procedures, functions and packages but those wrapped scripts can be unwrapped by using third party tools.*

*The scripts that we have developed analyzes all database sessions, and if they detect a DML or a DDL statement from an unauthorized user to procedure, function or package which should be protected then the execution of the statement is denied. Furthermore, these scripts do not allow a dba user to drop or disable the scripts themselves. In other words by managing sessions prior to the execution of an eventual statement from a dba user, we can prevent the execution of eventual statements which target our scripts.*

## 1. INTRODUCTION

The electronic information is simpler to reuse and modify and query, so the future is going towards to only digital data. Sometimes, we do not need to have a trace of changes that we made, and sometimes it is required to know the history of changes. This information can be stored in any database, but not all the database engines meet these requirements. The most important requirements vary to performance, hardware resources, security, user-friendliness, price or others.

The users, who have technical access to a lot of data, can use this information even if they do not have legal permissions to see or modify it. Sometimes they can benefit from these data, by modifying or reading it.

Oracle database, like any other database, database administrator can grant permissions on any database object to a database to user, so some users have rights to insert data, some others to modify and some other users to modify it, but the DBA user can modify, delete, insert/create, update data or database objects. Securing data stored in database involves not only using strong password or policy, but also adequate access controls over database objects to users.

Software developer try to make their products secure, try to hide business logic when is necessary by coding inside database using procedures, functions, packages, etc., and also try to secure the intellectual property. Oracle relational database management systems (8i, 9i, 10g and

11i versions) have a technique to wrap PL/SQL source code with either the DBMS_DDL subprograms or wrap utility.

Can we unwrap a PL/SQL source code block? Pete Finnigan investigated the PL/SQL wrapping process and published the way how to unwrap a PL/SQL source code. For example we can try to unwrap a PL/SQL source code online at http://www.codecrete.net/UnwrapIt.

## 2. ORACLE DATABASE SECURITY

The intent of this section is to give a basic understanding of the security capabilities of Oracle relational database management system. It is not the scope of this paper to explain all of the security features and options available in Oracle relational database management system. The Oracle RDMBS has too many security features which make it an excellent database system for any database application. Data integrity, confidentiality, and availability are well-protected with a properly designed Oracle database. [4]

### 2.1. Authentication

Oracle allows for various types of authentication. Oracle-based authentication allows for Oracle database accounts with user-ids and strong passwords which are encrypted with a modified DES algorithm for each database connection. Oracle passwords are stored in an encrypted format in the data dictionary [8]. Each session key is unique, which means the key is not re-used in any other session. Oracle also supports authentications based on the operating system's user accounts which are then passed on to Oracle RDBMS [4].

### 2.2. Profiles

Oracle makes use of profiles to allow the database administrator to place specific restrictions, rules and controls on a number of system resources, password usage lifetime and various Oracle products. These profiles can be named, defined, and then assigned to groups of users or to specific users. There are two types of profiles as bellow.

- **System resource profiles:** Those profiles can be used to put user limits on certain system resources such as memory, CPU time, the number of data blocks that can be read per session, the number of concurrent active sessions, idle time, and the maximum connection time for a user, etc. Also, they can be used to define and enforce password rules such as account lockout after a number of failed login attempts, password life, etc. Perhaps the most impressive feature of system resources profile is the ability to have password complexity checked using a custom PL/SQL function. The database administrator/developer can write a custom code for a complex PL/SQL function to define how a password must be, such as number of alpha, numeric and/or numeric characters, minimum length, and how different a new password must be from the previous one.[4][6][7]

- **Product profiles:** To prevent users from accessing specific commands or all commands in Oracle SQL, SQL/Plus, and PL/SQL. Use of this option allows the administrator to do such things as prevent user access to the operating system (SQL/PLUS HOST command), and to prevent unauthorized copying of data from one table to another (SQL/PLUS COPY command).[7]

### 2.3. Privileges

By default, Oracle relational database management system does not give any privileges to new users. New users must be given privileges before they create database connection and execute any database operation. Oracle users cannot do anything unless they have the specific privilege to do so. There are too many privileges in Oracle database management systems that can be given, around 100 in all. There are two different types of privileges available to be granted to any user. They are system and object privileges.

System privileges allow a user to execute DDL statements to create or manipulate objects, but do not give access read or write to actual database objects. System privileges allow a user to execute different DDL commands such as ALTER, CREATE, EXECUTE ANY PROCEDURE, and DELETE.

Object privileges are used to allow execute DML statements to access to some specific database object, such as a table or view or sequence. The database administrators can give users access to a chosen sub-set of columns or rows in a table, rather than the entire table.[4] Oracle also allows for the user of the GRANT privilege which allows a user to GRANT their privileges to another user or role for objects that they own as bellow.[7][9]

### 2.4. Roles

Roles are used to make easy the management process of assigning privileges to users. Roles are first created and then given different system and object privileges that can be assigned to users. Oracle database users can be given multiple roles even if they have the same privileges. It is much easier to create group of privileges that are organized into roles and then assign the role to one or more users. Roles also can be protected with passwords and the password protected roles require that a password be provided before activating it. The password feature is very useful in cases where a user needs access to data through a third party application but it is not desirable to give the user direct access to the data through the use of another third party tool like reporting tools, etc. The password can be supplied by the application, thus preventing the user to even need to know the password [4] [9].

Oracle has three different default roles which have some privileges assigned. The Connect Role allows user login and the ability to create tables, indexes, etc. The Resource Role is very similar to the pervious, but allows for more advanced rights (the creation of function and triggers and procedures). The DBA (Database Administrator) Role is granted all system privileges needed to administer the database. [6]

### 2.5. Protecting Data Integrity

Oracle relational database management system provides several features to ensure data integrity in different cases such system failure, human error, or attacks. These features include redo log files, rollback segments, and LogMiner.

All data changes are recorded in at least two redo log files that are maintained by Oracle. In the event of a system failure or data corruption, the last good backup and the redo log should be restored to bring the system back to the state it was before the corruption or data loss.

LogMiner is a SQL-based log file analyser utility and can be used to analyze the redo log files and rollback segments. [9]

## 3. PL/SQL SOURCE CODE BLOCKS AND WRAPPING PROCESS

PL/SQL stands for Procedural Language/SQL. PL/SQL extends SQL by adding constructs found in procedural languages, resulting in a structural language that is more powerful than SQL. The basic unit in PL/SQL is a block. All PL/SQL programs are made up of blocks, which can be nested within each other. Typically, each block performs a logical action in the program. A block has the following structure:

```
DECLARE

   /* Declarative section: variables, types,
      and local subprograms. */

BEGIN

   /* Executable section: procedural and
      SQL statements go here. */

   /* This is the only section of the block
      that is required. */

    EXCEPTION

      /* Exception handling section:
            error handling statements go here. */

END;
```

Only the executable section is required. The other sections are optional. The only SQL statements allowed in a PL/SQL program are SELECT, INSERT, UPDATE, DELETE and several other data manipulation statements plus some transaction control. Data definition statements like CREATE, DROP, or ALTER are not allowed. The executable section also contains constructs such as assignments, branches, loops, procedure calls, and triggers, which are all described below (except triggers). PL/SQL is not case sensitive. C style comments (/* ... */) may be used.

A stored procedure, a function, a trigger or a package is a subroutine available to applications that access a relational database system. They are actually stored in the database data dictionary.

Typical uses include data validation (integrated into the database) or access control mechanisms. Furthermore, they can consolidate and centralize logic that was originally implemented in applications. Extensive or complex processing that requires execution of several SQL statements is moved into Pl/SQL code blocks, and all applications call the procedures.

Benefits from pl/sql objects (procedure, function, trigger and package) stored in database:

- **Overhead:** Because stored statements are stored directly in the database, they may remove all or part of the compilation overhead that is typically required in situations where software applications send inline SQL statements to a database. In addition, while they avoid some overhead, pre-compiled SQL statements add to the complexity of creating an optimal execution plan because not all arguments of the SQL statement are supplied at compile time [16].

- **Avoidance of network traffic:** A major advantage with stored pl/sql object is that they can run directly within the database engine. In a production system, this typically means that the procedures run entirely on a specialized database server, which has direct access to the data being accessed. The benefit here is that network communication costs can be avoided completely. This becomes particularly important for complex series of SQL statements. [16]

- **Encapsulation of business logic:** Database stored PL/SQL objects allows software developers to embed business logic in the database. This method can reduce the need to encode the logic elsewhere in any client applications. The database system with the help of stored procedures can ensure data integrity and consistency.

- **Delegation of access-rights:** In many systems, stored PL/SQL objects can be granted access rights to the database that users who execute those procedures do not directly have.

- **Protection from SQL injection attacks:** Database stored PL/SQL objects can be used to protect against injection attacks. Parameters used during function or a procedure execution will be treated as data even if they are inserts SQL commands from an attacker.

### 3.1. Wrapping Process

Wrapping is the process of hiding PL/SQL source code. Wrapping helps developers to protect their source code from any user who can benefit or might misuse it. Wrapping a package causes the code to be obfuscated, that is to say made unreadable by programmers, so that the code cannot be readily stolen, or looked through by hackers for weaknesses to exploit, thus protecting the intellectual property of the programming company.

However wrapped packages can still be used by others such as the accounting firm, despite the code being unreadable, so the programming company has guarded their secrets and still enabled the accounting firm to use their product.

Oracle relational database management system offers two different ways to wrap PL/SQL source code procedure, function or package with either the wrap utility or DBMS_DDL subprograms.

**Wrap utility:** The wrap utility is an external application which processes an input SQL file which contains different statements and wraps the PL/SQL units in it, such as a function, procedure, package, type specification, or type body. This utility does not wrap PL/SQL code in anonymous blocks or triggers or DML statements.

To run the wrap utility, the wrap command can be used via operating system prompt using the following syntax (with no spaces around the equal signs):

```
wrap iname=input_file [ oname=output_file ]
```

input_file is the name of a file containing SQL statements, that you typically run using SQL*Plus. If you omit the file extension, an extension of .sql is assumed.

**DBMS_DDL subprograms**: The DBMS_DDL package contains procedures for wrapping only a single PL/SQL unit, such as a function, procedure, package, type specification, or type body. These overloaded subprograms provide a mechanism for wrapping dynamically generated PL/SQL units. The DBMS_DDL package contains different object like the WRAP functions and the CREATE_WRAPPED procedures. The CREATE_WRAPPED procedure wraps the text and creates the PL/SQL unit.

In the following PL/SQL code is used to dynamically create and wrap a package in a database.

```
DECLARE
```

```plsql
   package_text VARCHAR2(32767);
   -- Text for creating package spec & body

   -- function generate_spec generates package specification
   FUNCTION generate_spec (pkgname VARCHAR2)
                     RETURN   VARCHAR2 AS

   BEGIN

      RETURN 'CREATE PACKAGE ' || pkgname || ' AS

         PROCEDURE raise_salary (emp_id NUMBER,
                                 amount NUMBER);

         PROCEDURE fire_employee (emp_id NUMBER);

         END ' || pkgname || ';';

   END generate_spec;

   FUNCTION generate_body (pkgname VARCHAR2)
                              RETURN VARCHAR2 AS
   BEGIN

      RETURN 'CREATE PACKAGE BODY ' || pkgname || ' AS
         PROCEDURE raise_salary (emp_id NUMBER,
                                 amount NUMBER) IS
         BEGIN

            UPDATE employees
            SET salary = salary + amount
            WHERE employee_id = emp_id;

         END raise_salary;

         PROCEDURE fire_employee (emp_id NUMBER) IS
         BEGIN

             DELETE FROM employees WHERE employee_id = emp_id;

         END fire_employee;

      END ' || pkgname || ';';

   END generate_body;

BEGIN

   -- Generate package spec
   package_text := generate_spec('emp_actions')

   -- Create wrapped package spec
   DBMS_DDL.CREATE_WRAPPED(package_text);

   -- Generate package body
   package_text := generate_body('emp_actions');

   -- Create wrapped package body
```

```
        DBMS_DDL.CREATE_WRAPPED(package_text);

    END;
    /

    -- Invoke procedure from wrapped package
    CALL emp_actions.raise_salary(120, 100);
```

When we check the static data dictionary views *_SOURCE, the source is wrapped, or hidden, so that others cannot view the code details. For example:

```
    SELECT text
    FROM USER_SOURCE
    WHERE name = 'EMP_ACTIONS';
```

The resulting output is similar to the following:

```
    TEXT
    --------------------------------------------------
    PACKAGE emp_actions WRAPPED
    a000000
    1f
    abcd
    ...
```

Wrapped source files can be modified, copied, moved, and processed by SQL*Plus and the import and export utilities, but they hidden and cannot be visible through the static data dictionary views *_SOURCE.

### 3.1.1 Limitations of Wrapping

- Wrapping process is not recommended for hiding passwords or table names because is not a secure method. This process prevents most users from examining the source code of a PL/SQL unit, but might not stop all of them.

- Wrapping does not hide the source code for database triggers. To prevent users from examining the source code workings of a trigger, a one-line trigger that invokes a wrapped subprogram is necessary.

- Wrapping process does not detect syntax or semantic errors (for example, nonexistent tables or views) it detects only tokenization errors (for example, runaway strings). Syntax or semantic errors are detected during PL/SQL code compilation or during execution.

- Wrapped PL/SQL units are not downward-compatible between Oracle Database releases they are only upward-compatible. For example, a file processed by the V8i wrap utility can be used into a V9i Oracle Database, but a file processed by the V9i wrap utility cannot be used into a V8i Oracle Database.

### 3.2 Unwrapping Process

In old versions of Oracle we can almost easily deduce the original source code of PL/SQL package from the wrapped code produced. The symbol table is visible, so we can understand the purpose of the procedures (because we see variable names in symbol table), we can find out encryption algorithms that used. We also can modify the source code working straight with the wrapped code.

In version 9i of Oracle almost the same information of the source code we can get in 9i analyzing DIANA code.

In version 10g and 11i of Oracle Finnigan claims that unwrapping is almost the same as for 9i version, though some it is more difficult now because new wrap mechanism is provided, the symbol table is no longer visible, used base64 encryption, but IDL$ (DIANA) tables still contain DIANA m-code [5].

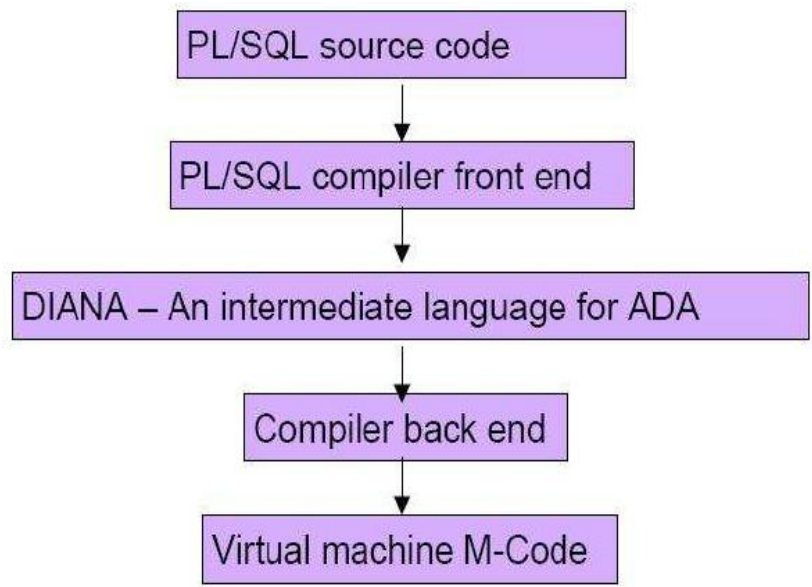

Figure 1. PL/SQL language compilation structure.

To unwrap a PL/SQL exist too many free tools or online websites like http://www.codecrete.net/UnwrapIt .

It is beyond the scope of this paper to cover all of the unwrapping process.

## 4. PROTECTING PL/SQL OBJECTS BY DISABLING DML/DDL STATEMENTS TO ANY USER

To protect PL/SQL objects is necessary to disable DDL statements like create, replace and some DML statements oriented to the static data dictionary views *_SOURCE. To prevent execution of any DML/DDL statements on some object we must monitor every user session and analyze every SQL statements before their execution.

We developed some scripts which can prevent any unauthorized user to execute DML/DDL statements over any database object. Those scripts monitor every user sessions and if they detect any DML/DDL statements over objects, which are protected by scripts, then scripts immediately, kill this user session and do not allow execution of those DML statements.

The scripts also analyze DDL statements, because if a user with permission to disable triggers, to drop package, to modify procedure, functions, and etc, can disable our scripts too. By analyzing DDL statements, scripts can protect their self from any unauthorized DDL statements. So if a user is trying to execute DDL statements over our scripts, this session will be killed immediately [9].

All the DML and DDL statements are checked, and if statement is not allowed to be executed, immediately their session will be killed [9].

The package of those scripts also contains some tables and procedures.

Table 1. The package's objects.

| Object name | Type | Description |
|---|---|---|
| set_security | Procedure | This procedure enables and disables protection. Procedure password is required. |
| set_password | Procedure | This procedure change password. Old and new password a required. |
| reset_password | Procedure | This procedure sends an email to security officer with new password. |
| add_object | Procedure | This procedure defines a new object to be protected. Object owner, object type and object name are required. |
| remove_object | Procedure | This procedure removes an object from protection list. Object owner, object type and object name are required. |
| grant_permission | Procedure | This procedure grants permission on a protected object to a database user. Object owner, object type and object name, and user name are required. Start date, end date, start hour and end hour are optional. |
| revoke_permission | Procedure | This procedure revokes permission a database user. Object owner, object type and object name, and user name are required. |
| exp_killed_session | Table | This procedure exports all killed sessions. Start date and end date are optional. |
| security_object | Table | This table contains information about objects which will be protected |
| user_permission | Table | This table contains information about users which have permission over protected objects |
| p_config | Table | This table contains information about configuration, password, etc. data are encrypted |
| killed_sessions | Table | This table contains information about killed sessions |
| ddl_log | Table | This table contains information about every DDL statement |

| | | executed by any database user |
|--|--|--|

# 5 CONCLUSIONS & FUTURE WORK

Our scripts can disable DML/DDL statements of an unauthorized user even he has DBA role, but we must monitor all sessions, so the database performance decrease but it does not influence too much. Our tests were based on a database schema which includes about 300 tables and some table contains more 50,000,000 million records. After the test we did not find a difference on a query time before the implementation of scripts and after it.

As described above we can stop execution of DDL statements oriented to the static data dictionary views *_SOURCE, but this prevent DBA to access its PL/SQL objects, to solve this problem session monitoring part will allow query when their results do not contain any part from of protected PL/SQL objects. This part must be modified in future to allow execution of DML statements oriented to the static data dictionary views *_SOURCE but have to modify the result where protected objects will be invisible to user request.

Even if DBA cannot execute DML/DDL statements on protected objects or DDL statements over our package he still can manipulate the data by generating some triggers to an authorized user. To eliminate this problem we insert every DDL statements into ddl_log table which is part of our package.

Package must be installed to a user with sysdba role because it uses some necessary packages.

During protection of objects, the scripts kill untrusted sessions but sometime not all the SQL statements on those sessions must be killed.

**Authors**

**Hakik PACI:** PHD Student, Polytechnic University of Tirana (2011). I have a strong background in Relational Database Systems, Database Administration, and Performance improvement and in Software Design.